\documentclass[preprint,aps,  nofootinbib]{revtex4}
\usepackage{amssymb}
\usepackage{amsmath}
\usepackage{graphicx}

\newcommand\beq{ \begin{eqnarray} }
\newcommand\eeq{ \end{eqnarray} }

\begin{document}

\title{QCD Viscosity to Entropy Density Ratio in the Hadronic Phase}
\author{Jiunn-Wei Chen}
\author{Yen-Han Li}
\author{Yen-Fu Liu}
\author{Eiji Nakano}
\affiliation{Department of Physics and Center for Theoretical Sciences, National Taiwan
University, Taipei 10617, Taiwan}

\begin{abstract}
Shear viscosity $\eta $ of QCD in the hadronic phase is computed by the
coupled Boltzmann equations of pions and nucleons in low temperatures and
low baryon number densities. The $\eta $ to entropy density ratio $\eta /s$
maps out the nuclear gas-liquid phase transition by forming a valley tracing
the phase transition line in the temperature-chemical potential plane. When
the phase transition turns into a crossover, the $\eta /s$ valley gradually
disappears. We suspect the general feature for a first-order phase
transition is that $\eta /s$ has a discontinuity in the bottom of the $\eta
/s$ valley. The discontinuity coincides with the phase transition line and
ends at the critical point. Beyond the critical point, a smooth $\eta /s$
valley is seen. However, the valley could disappear further away from the
critical point. The $\eta /s$ measurements might provide an alternative to
identify the critical points.
\end{abstract}

\maketitle


\section{Introduction}

Shear viscosity $\eta $ is a transport coefficient which has recently been
attracting lots of attention. It characterizes how strongly particles
interact and move collectively in a many-body system. In general, the
stronger the interparticle interaction, the smaller the shear viscosity. It
is conjectured \cite{KOVT1} that, no matter how strong the interparticle
interaction is, the shear viscosity to entropy density $s$ ratio has a
minimum bound $1/4\pi $. I.e. $\eta /s$ $\geq 1/4\pi $ in any system. The
bound was motivated by the uncertainty principle and the observation that $%
\eta /s$ $=1/4\pi $ for a large class of strongly interacting quantum field
theories whose dual descriptions in string theory involve black holes in
anti-de Sitter space \cite%
{Policastro:2001yc,Policastro:2002se,Herzog:2002fn,Buchel:2003tz}. In Ref. 
\cite{KOVT1}, supporting evidence of the conjecture was given for matters
like H$_{2}$O, He and N. Their $\eta /s$ curves reach their minima near the
gas-liquid phase transitions with the bound well satisfied. Recently, $\eta
/s$ close to the minimum bound was found in relativistic heavy ion
collisions (RHIC) \cite{RHIC,Molnar:2001ux,Teaney:2003pb} (and in lattice
simulations of a gluon plasma \cite{Nakamura:2004sy}) just above the
deconfinement temperature $T_{c}$($\sim 170$ MeV at zero baryon density \cite%
{KL04}). This suggests the quark gluon plasma (QGP) is strongly interacting
at this temperature, which is quite different from the traditional picture
of weakly interacting QGP. \footnote{%
See also \cite{Hatsuda03,Datta03,Umeda02}. For discussions of the possible
microscopic stucture of such a state, see \cite%
{Shuryak:2004tx,Koch:2005vg,Liao:2005pa,GerryEd,GerryRho}} (However, see
Ref. \cite{Asakawa:2006tc} for a\ different interpretation.) Also, $\eta /s$
close to the minimum bound was found in cold fermionic atoms in the infinite
scattering length limit \cite{Schafer:2007pr}. A relation between $\eta /s$
and the jet quenching parameter in QGP was proposed in Ref. \cite%
{Majumder:2007zh}.

In Refs. \cite{Csernai:2006zz} and \cite{Chen:2006ig}, it was found that $%
\eta /s$ of QCD in the confinement and deconfinement phases is qualitatively
different. When $T<T_{c}$ (the confinement phase), $\eta /s$ is
monotonically decreasing in $T$ because the system is dominated by goldstone
bosons which interact more weakly at lower $T.$ When $T>T_{c}$ \ (the
deconfinement phase), $\eta /s$ is monotonically increasing in $T$ because
the interaction between quarks and gluons is weaker at higher $T$ due to
asymptotic freedom. It makes perfect sense to have the phase transition from
the point of view of preserving the $\eta /s$ minimum bound. This is because
if the qualitative behavior of $\eta /s$ is not changed by a phase
transition (or crossover), then the bound could be violated. One concludes
that the minimum or valley of the $\eta /s$ curve lies in the vicinity of $%
T_{c}$ (the extrapolation of the low(high) temperature $\eta /s$ curve sets
a upper(lower) bound on $T_{c}$) \cite{Chen:2006ig}. This behavior is also
seen in the H$_{2}$O, He and N systems. It was further noticed that below
the critical pressure, a cusp appears at the minimum of $\eta /s$\ which
coincides with the critical temperatures \cite{Csernai:2006zz}.

In Ref. \cite{Cohen:2007qr}, it is argued that a universal minimum bound on $%
\eta /s$ should not exist. The counterexample given is a system of mesons
made by heavy quarks and light antiquarks. Since $s$ scales linearly with
the number of heavy quark flavor $N_{f}$ and $\eta $ is insensitive to $%
N_{f} $, the $\eta /s$ $\geq 1/4\pi $ bound could be violated in some
special large $N_{f}$ limit. However, this system is metastable. As far as
the qualitative relation between $T_{c}$ and the valley of $\eta /s$ is
concerned, it does not matter the minimum bound of $\eta /s$ is $1/4\pi $ or 
$0$. As long as there is a lower bound, the monotonic behavior of $\eta /s$
will be affected by the bound.

In this manuscript, we extend the discussion of the $\eta /s$ of QCD in the
confinement phase at zero baryon chemical potential $\mu $ \cite{Chen:2006ig}%
\ to finite $\mu $ and study its relation to the QCD phase diagram. String
theory methods give $\eta /s$ $=1/4\pi $ for $\mathcal{N}=4$ supersymmetric
theories with finite R-charge density, suggesting the minimum bound is
independent of $\mu $. For QCD, the fermion sign problem (the fermion
determinant is not positive definite) makes the current lattice QCD methods
inapplicable in the low $T$ and finite $\mu $ regime. An alternative is to
use effective field theory (EFT). Reliable results using EFT in hadronic
degrees of freedom can be obtained when both $T$ and $\mu $ are small. At
higher $\mu $ (with $\left\vert k_{F}a\right\vert \gg 1$, where $k_{F}$
denotes the fermi momentum and $a$ is the nucleon-nucleon scattering length)
the problem becomes non-perturbative in coupling and mean-field treatments
are not sufficient. (It is essentially the same type of problem as in cold
fermionic atoms near the infinite scattering length limit.) Non-perturbative
computations of the EFT on the lattice is free from the fermion sign problem
at the leading\ order (LO) with only non-derivative contact interactions 
\cite{Chen:2003vy}. But this theory is suitable only in low $T$ and low
density systems. For the nuclear matter problem, the inclusion of one pion
exchange will re-introduce the sign problem. However, the sign problem is
claimed to be mild and lattice simulations are still possible \cite%
{Borasoy:2006qn}. The computation of $\eta $ using lattice nuclear EFT has
not been carried out before. Although $\eta $ is associated with real time
response to perturbations, it can be reconstructed through the\ spectral
function computed on the Euclidean lattice \cite%
{Nakamura:2004sy,Karsch:1986cq}.

As an exploratory work, we compute $\eta $ using coupled Boltzmann equations
for a system of pion $\pi $ and nucleon $N$ while the entropy $s$ is
computed only for free particles. This approach will not give accurate $\eta
/s$ in the regime dominated by near threshold $NN$ interaction. However, for
most of the regime we are exploring, our result should be robust.


\begin{figure}[tbp]
\begin{center}
\includegraphics[height=6cm]{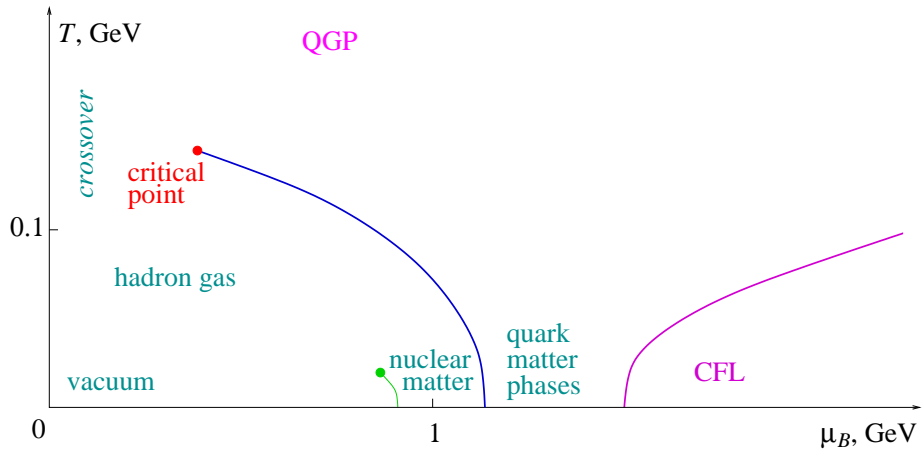}
\end{center}
\caption{A semi-quantitative sketch of the QCD phase diagram \protect\cite%
{Stephanov:2007fk} courtesy of \ M. Stephanov. }
\end{figure}


\section{Linearized Boltzmann Equation for Low Energy QCD}

We are interested in the hadronic phase of QCD with non-zero baryon-number
chemical potential $\mu $. For reference, a schematic QCD phase diagram from
a recent review \cite{Stephanov:2007fk} is shown in Fig. 1. The detail
structures of the quark matter phases are still unclear. When $\mu \ll m_{N}$
($=938$ MeV, the nucleon mass) the nucleon population is exponentially
suppressed. The dominant degrees of freedom are the lightest hadrons---the
pions. The pion mass $m_{\pi }$($=139$ MeV) is much lighter than the mass of
the next lightest hadron---the kaon whose mass is $495$ MeV. Given that $%
T_{c}$ is only $\lesssim 170$ MeV, it is sufficient to just consider the
pions in the calculation of thermodynamical quantities and transport
coefficients for $T\ll T_{c}$. When $m_{N}-\mu <m_{\pi }$, the nucleon
population is no longer suppressed compared with the pion. We will limit
ourselves to the low $T$ and low $\mu $ region where only $\pi $ and $N$ are
important degrees of freedom.

The shear viscosity of a system is defined by the Kubo formula%
\begin{equation}
\eta =-\frac{1}{5}\int_{-\infty }^{0}\mathrm{d}t^{\prime }\int_{-\infty
}^{t^{\prime }}\mathrm{d}t\int \mathrm{d}x^{3}\langle \left[
T^{ij}(0),T^{ij}(\mathbf{x},t)\right] \rangle \ ,
\end{equation}%
with $T^{ij}$ the spacial part of the off-diagonal energy momentum tensor.
The Kubo formula involves an infinite number of diagrams at the LO even in
the weak coupling $\phi ^{4}$ theory \cite{Jeon}. However, it is proven that
the summation of LO diagrams in a weak coupling $\phi ^{4}$ theory is
equivalent to solving the linearized Boltzmann equation with temperature
dependent particle masses and scattering amplitudes \cite{Jeon}. We will
assume the equivalence between the Kubo formula and the Boltzmann equation
still hold in our $\pi N$ system. Later we will check whether the mean free
path is still much larger than the range of interaction. This is a
requirement to apply the Boltzmann equation which makes use of
semi-classical descriptions of particles with definite position, energy and
momentum except during brief collisions.

In the Boltzmann equation of our $\pi N$ system, the evolution of the
isospin averaged $\pi $ and $N$ distribution functions $f^{\pi ,N}=f^{\pi
,N}(\mathbf{x},\mathbf{p},t)\equiv f_{p}^{\pi ,N}(x)$ (functions of space,
time and momentum) are caused by interparticle $\left( \pi \pi \text{, }\pi N%
\text{ and }NN\right) $ collisions 
\begin{eqnarray}
\frac{p^{\mu }}{E_{p}^{\pi }}\partial _{\mu }f_{p}^{\pi }(x) &=&\frac{g_{\pi
}}{2}\int_{123}d\Gamma _{12;3p}^{\pi \pi }\left\{ f_{1}^{\pi }f_{2}^{\pi
}F_{3}^{\pi }F_{p}^{\pi }-F_{1}^{\pi }F_{2}^{\pi }f_{3}^{\pi }f_{p}^{\pi
}\right\}  \notag \\
&&+g_{N}\int_{123}d\Gamma _{12;3p}^{\pi N}\left\{ f_{1}^{N}f_{2}^{\pi
}F_{3}^{N}F_{p}^{\pi }-F_{1}^{N}F_{2}^{\pi }f_{3}^{N}f_{p}^{\pi }\right\} \ ,
\label{B1} \\
\frac{p^{\mu }}{E_{p}^{N}}\partial _{\mu }f_{p}^{N}(x) &=&\frac{g_{N}}{2}%
\int_{123}d\Gamma _{12;3p}^{NN}\left\{
f_{1}^{N}f_{2}^{N}F_{3}^{N}F_{p}^{N}-F_{1}^{N}F_{2}^{N}f_{3}^{N}f_{p}^{N}%
\right\}  \notag \\
&&+g_{\pi }\int_{123}d\Gamma _{12;3p}^{\pi N}\left\{ f_{1}^{\pi
}f_{2}^{N}F_{3}^{\pi }F_{p}^{N}-F_{1}^{\pi }F_{2}^{N}f_{3}^{\pi
}f_{p}^{N}\right\} \ ,  \label{B2}
\end{eqnarray}%
where $F_{i}^{\pi (N)}\equiv 1\pm f_{i}^{\pi (N)}$, $E_{p}^{\pi (N)}=\sqrt{%
\mathbf{p}^{2}+m_{\pi (N)}^{2}}$, and the spin and isospin degeneracy
factors $g_{\pi }=3$ and $g_{N}=4$. 
\begin{equation}
d\Gamma _{12;3p}^{\pi N}\equiv |\mathcal{T}_{\pi N}|^{2}\frac{(2\pi
)^{4}\delta ^{4}(k_{1}+k_{2}-k_{3}-p)}{2^{4}E_{1}^{N}E_{2}^{\pi
}E_{3}^{N}E_{p}^{\pi }}\prod_{i=1}^{3}\frac{d^{3}\mathbf{k}_{i}}{(2\pi )^{3}}%
\ ,  \label{piN}
\end{equation}%
where $\mathcal{T}_{\pi N}$ is the $\pi N$ scattering amplitude with momenta 
$1,2\rightarrow 3,p$. The $\pi \pi $ and $NN$ weighted integration measures
are given analogously. We use the LO chiral perturbation theory ($\chi $PT) 
\cite{ChPT,GL,HBChPT,BKM} result for the isospin averaged $\pi \pi $
scattering amplitude in terms of Mandelstam variables ($s,t$, and $u$) 
\begin{equation}
|\mathcal{T}_{\pi \pi }|^{2}=\frac{1}{9f_{\pi }^{4}}\left[ 21m_{\pi
}^{4}+9s^{2}-24m_{\pi }^{2}s+3(t-u)^{2}\right] \ ,
\end{equation}%
where $f_{\pi }=93$ MeV. The isospin averaged $NN$ scattering amplitude are
described by effective range expansion \cite{Bethe}. In the center of mass
(CM) frame 
\begin{eqnarray}
|\mathcal{T}_{NN}|^{2} &=&3(4\pi m_{N})^{2}\left[ |-\frac{1}{a_{1}}+\frac{1}{%
2}r_{1}p^{2}-ip|^{-2}\right.  \notag \\
&+&\left. |-\frac{1}{a_{3}}+\frac{1}{2}r_{3}p^{2}-ip|^{-2}\right] \ ,
\end{eqnarray}%
where $p$ is the magnitude of nucleon momentum in the CM frame and $a_{1(3)}$
and $r_{1(3)}$ are the spin singlet(triplet) scattering length and effective
range, respectively. $a_{1}=-17.9$ \footnote{%
We have used the isospin averaged scattering length defined as $%
a_{1}^{2}=\left( a_{nn}^{2}+a_{np}^{2}+a_{pp}^{2}\right) /3$.}, $a_{3}=5.42$%
, $r_{1}=2.77$, and $r_{3}=1.76$, all in units of fm. Note that near the
threshold ($p=0$), $|\mathcal{T}_{NN}|^{2}$ is proportional to $%
a_{1}^{2}+a_{3}^{2}$ which is greatly enhanced by the large scattering
lengths. The interaction is smaller away from the threshold$.$

The $\pi N$ scattering [$\pi (q_{1})\,N(p_{1})$ $\rightarrow $ $\pi
(q_{2})\,N(p_{2})$ ] amplitude is also given by the LO $\chi $PT \cite%
{Fettes:1998ud}. In the CM frame 
\begin{eqnarray}
|\mathcal{T}_{\pi N}|^{2} &=&\frac{1}{2}\left( E_{N}+m_{N}\right) ^{2}\left[
2|g_{-}|^{2}+q^{\,2}\sin ^{2}\theta |h_{+}|^{2}\right] \ ,  \notag \\
g_{-} &=&-\frac{g_{A}^{2}}{f_{\pi }^{2}}\frac{1}{4\omega }\left( 2\omega
^{2}-2m_{\pi }^{2}+t\right) +\frac{\omega }{2f_{\pi }^{2}}\ ,  \notag \\
h_{+} &=&-\frac{g_{A}^{2}}{f_{\pi }^{2}}\frac{1}{2\omega }\ ,
\end{eqnarray}%
where $g_{A}=1.26$ is the $\pi N$ coupling constant, $E_{N}$ is the nucleon
energy, $q$ is the magnitude of pion momentum, $\theta $ is the angle
between $q_{1}$ and $q_{2}$, and $\omega $ is the pion energy. The thermal
corrections for $\pi \pi $, $\pi N$ scattering amplitudes and particle
masses are higher order in $\chi $PT. The $T$ dependence in $NN$ scattering
amplitude is also small at low $T.$ Because the thermal correction for the
inverse scattering length scales as $T$ which is much smaller than the
thermal momentum $\sim \sqrt{m_{N}T}$, it can be neglected.

In local thermal equilibrium, the distribution functions are $\overline{f}%
_{p}^{\pi }(x)=\left( e^{\beta (x)V_{\mu }(x)p^{\mu }}-1\right) ^{-1}$ and $%
\overline{f}_{p}^{N}(x)=\left( e^{\beta (x)V_{\mu }(x)\tilde{p}^{\mu
}}+1\right) ^{-1}$, where $\beta (x)$ is the inverse temperature, $V^{\mu
}(x)$ is the four velocity of the fluid at the space-time point $x$ and $%
\tilde{p}^{\mu }=(E_{p}^{N}-\mu ,\mathbf{p})$ in the $\mathbf{V}(x)=0$
frame. A small deviation of $f_{p}$ from local equilibrium can be
parametrized as 
\begin{equation}
f_{p}^{l}(x)=\overline{f}_{p}^{l}(x)\left[ 1-\overline{F}_{p}^{l}(x)\chi
_{p}^{l}(x)\right] \ ,\ \quad l=\pi ,N\ ,
\end{equation}%
with $\overline{F}_{i}^{\pi (N)}\equiv 1\pm \overline{f}_{i}^{\pi (N)}$. The
energy momentum tensor is 
\begin{equation}
T_{\mu \nu }(x)=\int \frac{\mathrm{d}^{3}\mathbf{p}}{(2\pi )^{3}}p_{\mu
}p_{\nu }\left[ \frac{g_{\pi }f_{p}^{\pi }(x)}{E_{p}^{\pi }}+\frac{%
g_{N}f_{p}^{N}(x)}{E_{p}^{N}}\right] \ .  \label{dT}
\end{equation}%
We will choose the frame with zero fluid velocity $\mathbf{V}(x)=0$ at the
point $x$. This implies $\partial _{\nu }V^{0}=0$ after taking a derivative
on $V_{\mu }(x)V^{\mu }(x)=1$. Furthermore, the conservation law at
equilibrium $\partial _{\mu }T^{\mu \nu }|_{\chi _{p}=0}=0$ allows us to
replace $\partial _{t}\beta (x)$ and $\partial _{t}\mathbf{V}(x)$ by terms
proportional to $\nabla \cdot \mathbf{V}(x)$ and $\mathbf{\nabla }\beta (x)$%
. Thus, to the first order in a derivative expansion, $\chi _{p}^{l}(x)$ can
be parametrized as 
\begin{eqnarray}
&&\frac{\chi _{p}^{l}(x)}{\beta (x)}=A^{l}(p)\nabla \cdot \mathbf{V}%
(x)+B_{ij}^{l}(p)\left( \frac{\nabla _{i}V_{j}(x)+\nabla _{j}V_{i}(x)}{2}-%
\frac{\delta _{ij}}{3}\nabla \cdot \mathbf{V}(x)\right) \ ,\quad  \label{df1}
\\
&&B_{ij}^{l}(p)\equiv B^{l}(p)\left( \hat{p}_{i}\hat{p}_{j}-\frac{1}{3}%
\delta _{ij}\right) \ ,  \notag
\end{eqnarray}%
where $i$ and $j$ are spatial indexes. $A$ and $B$ are functions of $x$ and $%
p$. However, we have suppressed the $x$ dependence.

Substituting Eq. (\ref{df1}) into the Boltzmann equation Eq. (\ref{B1}), one
obtains a linearized equation for $B^{\pi }$:%
\begin{eqnarray}
&&\left( p_{i}p_{j}-\frac{1}{3}\delta _{ij}\mathbf{p}^{2}\right) =\frac{%
g_{\pi }E_{p}^{\pi }}{2}\int_{123}d\Gamma _{12;3p}^{\pi \pi }\overline{F}%
_{1}^{\pi }\overline{F}_{2}^{\pi }\overline{f}_{3}^{\pi }(\overline{F}%
_{p}^{\pi })^{-1}  \notag \\
&&\quad \quad \times \left[ B_{ij}^{\pi }(p)+B_{ij}^{\pi
}(k_{3})-B_{ij}^{\pi }(k_{2})-B_{ij}^{\pi }(k_{1})\right]  \notag \\
&&\quad \quad +g_{N}E_{p}^{\pi }\int_{123}d\Gamma _{12;3p}^{\pi N}\overline{F%
}_{1}^{\pi }\overline{F}_{2}^{N}\overline{f}_{3}^{N}(\overline{F}_{p}^{\pi
})^{-1}  \notag \\
&&\quad \quad \times \left[ B_{ij}^{\pi }(p)-B_{ij}^{\pi
}(k_{1})+B_{ij}^{N}(k_{3})-B_{ij}^{N}(k_{2})\right]  \notag \\
&\equiv &g_{\pi }G_{ij}^{\pi \pi }\left[ B^{\pi }\right] +g_{N}{G_{1}}%
_{ij}^{\pi N}\left[ B^{\pi }\right] +g_{N}{G_{2}}_{ij}^{\pi N}\left[ B^{N}%
\right] \ .  \label{LB1}
\end{eqnarray}%
The analogous equation for $B^{N}$ is obtained by replacing $\pi
\leftrightarrow N$ on the right-hand side of the above equation. There are
two other integral equations involving $A^{l}(p)\nabla \cdot \mathbf{V}(x)$
which are related to the bulk viscosity $\zeta $. They will not be treated
in this work.

In fluid dynamics the energy momentum tensor at equilibrium depends on
pressure $P(x)$ and energy density $\epsilon (x)$ as $T_{\mu \nu
}^{(0)}(x)=\left\{ P(x)+\epsilon (x)\right\} V_{\mu }(x)V_{\nu
}(x)-P(x)\delta _{\mu \nu }$. A small deviation away from equilibrium gives
additional contribution to $T_{\mu \nu }=T_{\mu \nu }^{(0)}+\delta T_{\mu
\nu }$ whose spatial components define the shear and bulk viscosity 
\begin{equation}
\delta T_{ij}=\zeta \delta _{ij}\nabla \cdot \mathbf{V}(x)-\eta \left(
\nabla _{i}V_{j}(x)+\nabla _{j}V_{i}(x)-\frac{2}{3}\delta _{ij}\nabla \cdot 
\mathbf{V}(x)\right) \ .\quad \quad
\end{equation}%
Comparing the above definition with Eqs.~(\ref{dT}) and (\ref{df1}), we
obtain%
\begin{eqnarray}
\eta &=&g_{\pi }L^{\pi }\left[ B^{\pi }\right] +g_{N}L^{N}\left[ B^{N}\right]
\ ,  \notag \\
L^{l}\left[ B^{l}\right] &=&\frac{\beta }{15}\int \frac{\mathrm{d}^{3}%
\mathbf{p}\,\mathbf{p}^{2}}{(2\pi )^{3}E_{p}^{k}}\overline{f}_{p}^{l}%
\overline{F}_{p}^{l}B^{l}(p)\ .  \label{eta}
\end{eqnarray}%
Now one sees immediately that if all the $\pi \pi $, $\pi N$ and $NN$
scattering cross sections are reduced by a factor $\lambda $, then Eq.~(\ref%
{LB1}) implies the $B^{l}$ functions will be $\lambda $ times larger. Then
by Eq.~(\ref{eta}), $\eta $ will be $\lambda $ times larger as well. This is
the non-perturbative feature of the Boltzmann equation. It gives a divergent 
$\eta $ for a non-interacting theory.

Contracting both sides of Eq.~(\ref{LB1}) by $\left( \hat{p}_{i}\hat{p}%
_{j}-\delta _{ij}/3\right) $ and applying it to Eq.~(\ref{eta}) yields 
\begin{eqnarray}
\eta &=&\frac{\beta }{10}\int \frac{\mathrm{d}^{3}\mathbf{p}}{(2\pi )^{3}}%
\frac{g_{\pi }}{E_{p}^{\pi }}\overline{f}_{p}^{\pi }\overline{F}_{p}^{\pi
}B_{ij}^{\pi }(p)  \notag \\
&&\times \left\{ g_{\pi }{G}_{ij}^{\pi \pi }\left[ B^{\pi }\right] +g_{N}{%
G_{1}}_{ij}^{\pi N}\left[ B^{\pi }\right] +g_{N}{G_{2}}_{ij}^{\pi N}\left[
B^{N}\right] \right\}  \notag \\
&&+\left[ \pi \leftrightarrow N\right]  \notag \\
&\equiv &g_{\pi }^{2}\langle B_{\pi }|{G}^{\pi \pi }\left[ B_{\pi }\right]
\rangle +g_{\pi }g_{N}\left\{ \langle B_{\pi }|{G_{1}}^{\pi N}\left[ B_{\pi }%
\right] \rangle +\langle B_{\pi }|{G_{2}}^{\pi N}\left[ B_{N}\right] \rangle
\right\}  \notag \\
&&+\left[ \pi \leftrightarrow N\right] \ .  \label{eta2}
\end{eqnarray}

To compute $\eta $, one can just solve $B^{\pi (N)}(p)$ from Eq. (\ref{LB1}%
). But here we follow the approach outlined in Ref. \cite{DOBA1,DOBA2} to
assume that $B^{\pi (N)}(p)$ is a\ smooth function which can be expanded
using a specific set of orthogonal polynomials: 
\begin{eqnarray}
B^{l}(p) &=&|\mathbf{p}|^{y}\sum_{r=0}^{\infty }b_{r}^{l}B_{l}^{(r)}(z(p))\ 
\notag \\
&\equiv &\sum_{r=0}^{\infty }b_{r}^{l}\widetilde{B}_{l}^{(r)}(z(p))\ ,\
\quad l=\pi ,N\ .  \label{BP1}
\end{eqnarray}%
where $B_{l}^{(r)}(z)$ is a polynomial up to $z^{r}$ and $b_{r}^{l}$ is its
coefficient. The overall factor $|\mathbf{p}|^{y}$ will be chosen by trial
and error to get the fastest convergence. We find that using $y=1.89$ and $%
z(p)=|\mathbf{p}|$, the series converges rather rapidly. The orthogonality
condition 
\begin{equation}
\frac{\beta }{15}\int \frac{\mathrm{d}^{3}\mathbf{p}}{(2\pi )^{3}}\frac{%
\left\vert \mathbf{p}\right\vert ^{2+y}}{E_{p}^{l}}\overline{f}_{p}^{l}%
\overline{F}_{p}^{l}B_{l}^{(r)}(z)B_{l}^{(s)}(z)=L_{l}^{(r)}\delta _{r,s}\ 
\label{BP2}
\end{equation}%
can be used to construct the $B_{l}^{(r)}(z)$ polynomials up to
normalization constants. For simplicity, we will choose 
\begin{equation}
B_{l}^{(0)}(z)=1\ .  \label{Con1}
\end{equation}

With this setup, Eqs.(\ref{eta}) and (\ref{Con1}) yield 
\begin{equation}
\eta =\sum_{r}\left[ g_{\pi }b_{r}^{\pi }L_{\pi
}^{(r)}+g_{N}b_{r}^{N}L_{N}^{(r)}\right] \delta _{0,r}\ ,
\end{equation}%
while Eq.(\ref{eta2}) yields 
\begin{eqnarray}
\eta &=&\sum_{r,s=0}^{\infty }g_{\pi }b_{r}^{\pi }\left\{ g_{\pi }b_{s}^{\pi
}\left\langle \widetilde{B}_{\pi }^{(r)}|{G}^{\pi \pi }\left[ \widetilde{B}%
_{\pi }^{(s)}\right] \right\rangle +g_{N}\left[ b_{s}^{\pi }\left\langle 
\widetilde{B}_{\pi }^{(r)}|{G_{1}}^{\pi N}\left[ \widetilde{B}_{\pi }^{(s)}%
\right] \right\rangle +b_{s}^{N}\left\langle \widetilde{B}_{\pi }^{(r)}|{%
G_{2}}^{\pi N}\left[ \widetilde{B}_{N}^{(s)}\right] \right\rangle \right]
\right\}  \notag \\
&&+\left[ \pi \leftrightarrow N\right] \ .
\end{eqnarray}

The $b_{r}^{\pi }$'s and $b_{r}^{N}$'s are functions of $\mu $ and $T$. In
general they are independent of each other. Thus the above two equations
yield 
\begin{equation}
\delta _{r,0}\left( 
\begin{array}{c}
L_{\pi }^{(0)} \\ 
L_{N}^{(0)}%
\end{array}%
\right) =\sum_{s}\left( 
\begin{array}{cc}
\pi \pi ^{rs} & \pi N^{rs} \\ 
N\pi ^{rs} & NN^{rs}%
\end{array}%
\right) \left( 
\begin{array}{c}
g_{\pi }b_{s}^{\pi } \\ 
g_{N}b_{s}^{N}%
\end{array}%
\right) \ ,  \label{ME2}
\end{equation}%
where the matrix elements are given by 
\begin{eqnarray}
\pi \pi ^{rs} &=&\langle \widetilde{B}_{\pi }^{(r)}|G^{\pi \pi }[\widetilde{B%
}_{\pi }^{(s)}]\rangle +\frac{g_{N}}{g_{\pi }}\langle \widetilde{B}_{\pi
}^{(r)}|G_{1}^{\pi N}[\widetilde{B}_{\pi }^{(s)}]\rangle \ ,  \notag \\
\pi N^{rs} &=&\langle \widetilde{B}_{\pi }^{(r)}|G_{2}^{\pi N}[\widetilde{B}%
_{N}^{(s)}]\rangle \ ,  \notag \\
NN^{rs} &=&\langle \widetilde{B}_{N}^{(r)}|G^{NN}[\widetilde{B}%
_{N}^{(s)}]\rangle +\frac{g_{\pi }}{g_{N}}\langle \widetilde{B}%
_{N}^{(r)}|G_{1}^{N\pi }[\widetilde{B}_{N}^{(s)}]\rangle \ ,  \notag \\
N\pi ^{rs} &=&\langle \widetilde{B}_{N}^{(r)}|G_{2}^{N\pi }[\widetilde{B}%
_{\pi }^{(s)}]\rangle \ .
\end{eqnarray}%
The matrix equation in Eq.~(\ref{ME2}) allows us to solve for $b_{s}$ and
obtain the shear viscosity. Since the expansion in Eq.~(\ref{BP1}) converges
rapidly, one does not need to keep many terms to solve for $\eta $. If only
the $s=0$ term is kept, then 
\begin{equation}
\eta \simeq \left( 
\begin{array}{c}
L_{\pi }^{(0)} \\ 
L_{N}^{(0)}%
\end{array}%
\right) ^{T}\left( 
\begin{array}{cc}
\pi \pi ^{00} & \pi N^{00} \\ 
N\pi ^{00} & NN^{00}%
\end{array}%
\right) ^{-1}\left( 
\begin{array}{c}
L_{\pi }^{(0)} \\ 
L_{N}^{(0)}%
\end{array}%
\right) \ .
\end{equation}

The computation of the entropy density $s$ is more straightforward since $s$%
, unlike $\eta $, does not diverge in a free theory. The contributions from $%
\pi \pi $ and $\pi N$ scattering are perturbative and are higher-order
effects in $\chi $PT. The $NN$ contribution is also perturbative at low $\mu 
$, but it becomes non-perturbative at $\mu \approx m_{N}$ and low $T$. In
this regime, the system is governed by near threshold $NN$ interaction. With
that limitation in mind, we compute $s$ as a free gas: 
\begin{equation}
s=-\beta ^{2}\frac{\partial }{\partial \beta }\frac{g_{\pi }\log {Z_{\pi }}%
+\,g_{N}\log {Z_{N}}}{\beta }\ ,  \label{S}
\end{equation}%
where the partition functions ${Z_{\pi (N)}}$ for free pions(nucleons) is 
\begin{equation}
\frac{\log {Z_{\pi (N)}}}{\beta }=-\frac{1}{\beta }\int \frac{\mathrm{d}^{3}%
\mathbf{p}}{(2\pi )^{3}}\log \left\{ 1\mp e^{-\beta \tilde{E}_{p}^{\pi
(N)}}\right\} \ ,
\end{equation}%
with $\tilde{E}_{p}^{\pi }\equiv E_{p}^{\pi }$ and $\tilde{E}_{p}^{N}\equiv
E_{p}^{N}-\mu $, up to temperature independent terms. \ 

\section{$\protect\eta /s$ and the QCD phase diagram}


\begin{figure}[tbp]
\begin{center}
\includegraphics[height=12cm]{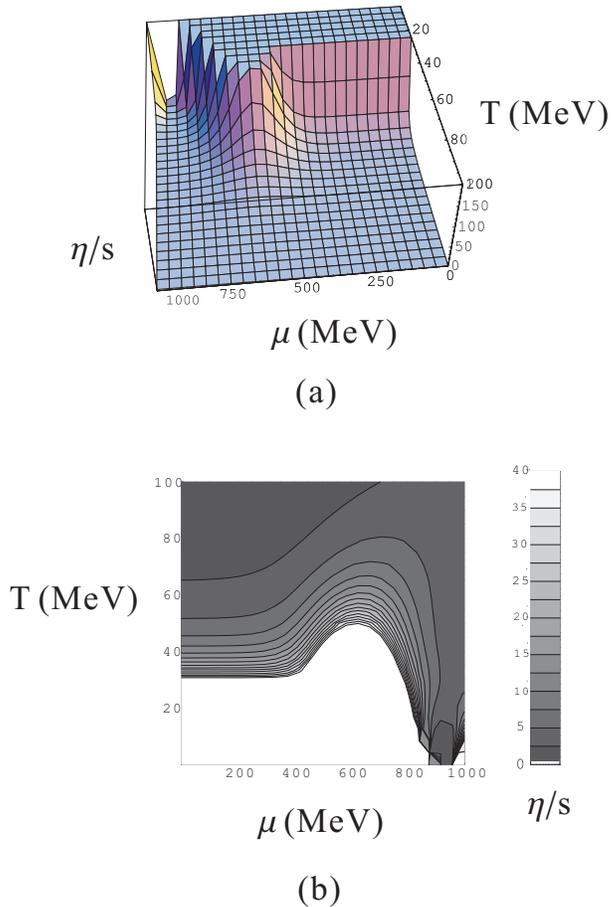}
\end{center}
\caption{$\protect\eta /s$ of QCD shown as a function of $T$ and $\protect%
\mu $ shown as a 3-D plot (a) and a contour plot (b). Note that the corner
of large $\protect\mu $ and large $T$ should be discarded since it is not in
the hadronic phase.}
\end{figure}


In Fig. 2, $\eta /s$ as a function of $T$ and $\mu $ is shown as a 3-D plot
and a contour plot. Note that at the corner of large $\mu $ and large $T$,
the system is no longer in the hadronic phase. Thus, the result should be
discarded there. In general $\eta /s$ is decreasing in $T$ except when $\mu
\simeq m_{N}$ and $T<30$ MeV. (This regime is blown up in Fig. 3 and will be
studied later.) There are some interesting structures at larger $\mu $, but $%
\eta /s$ is $\mu $ independent when $\mu <500$ MeV . This is because when $%
\mu \ll m_{N}$ the nucleons only exist through particle-antiparticle pair
creations, thus they are highly suppressed. The $\eta /s$ is determined by
the pion gas which is $\mu $ independent. Our result just reproduces the $%
\mu =0$ result of Ref. \cite{Chen:2006ig} (see \cite%
{DOBA1,DOBA2,Davesne,Prakash:1993kd} for earlier results) in this regime.

When $\mu >m_{N}-m_{\pi }=800$ MeV, the nucleon population is no longer
suppressed compared with the pion population and when $\mu \gtrsim m_{N}$
the nucleons become the dominant degrees of freedom. Numerically $\eta /s$
is dominated by the nucleon contributions when $\mu >800$ MeV. It is
decreasing in both $T$ and $\mu $ until $\mu \simeq m_{N}$. This is because $%
s$ is increasing in both $T$ and $\mu $ while $\eta $ is getting smaller at
higher $\mu $ (larger nucleon population) and lower $T$ (stronger
interaction, closer to the interaction threshold). 500-800 MeV in $\mu $ is
the transition between the $\pi $ and $N$ dominant regimes.


\begin{figure}[tbp]
\begin{center}
\includegraphics[height=12cm]{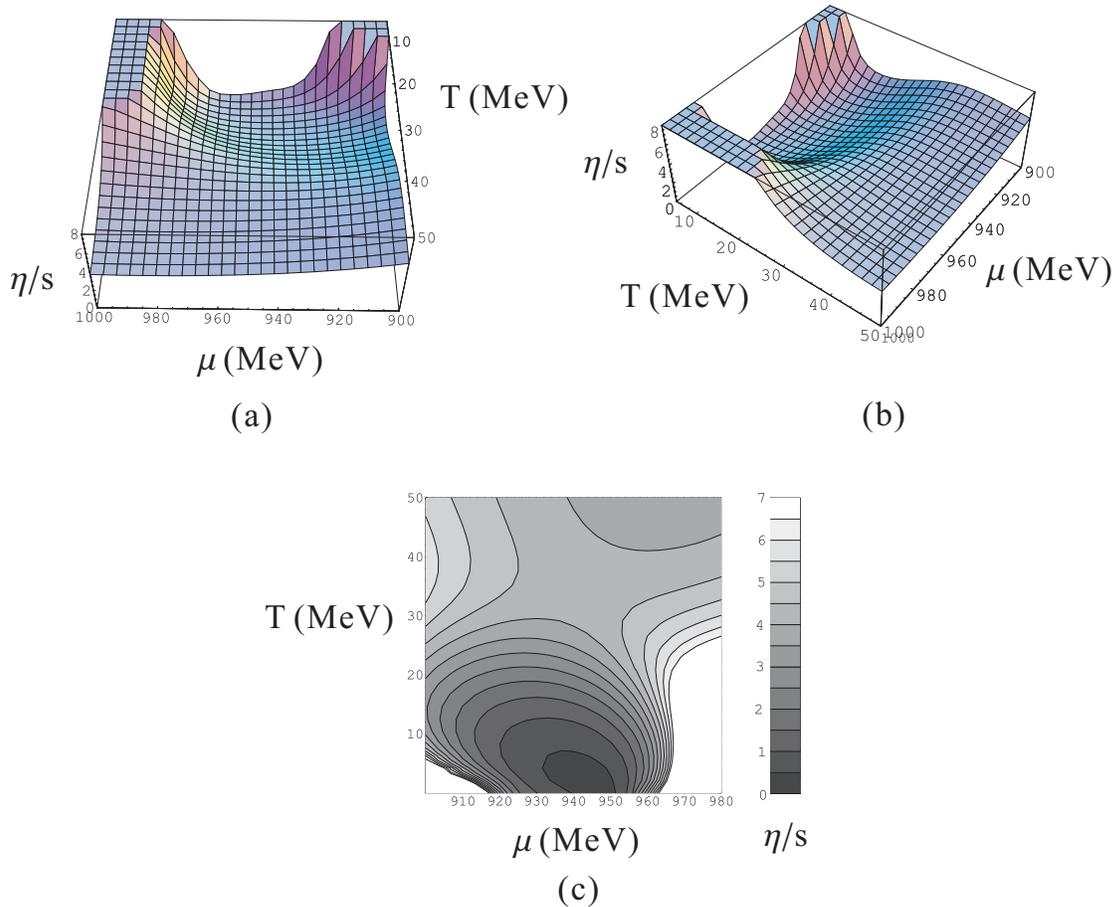}
\end{center}
\caption{$\protect\eta /s$ of QCD near the nuclear gas-liquid phase
transition shown as 3-D plots (a) and (b), viewed from different angles, and
a contour plot (c). The $\protect\eta /s$ maps out the nuclear gas-liquid
phase transition shown in Fig.1 by forming a valley tracing the nuclear
gas-liquid phase transition line in the $T$-$\protect\mu $ plane. When the
phase transition turns into a crossover at larger $T$, the valley also
gradually disappears at around 30 MeV. There should be a discontinuity that
looks like a fault in the bottom of the $\protect\eta /s$ valley that is not
seen in our approximation. The fault would lie on top of the phase
transition line and end at the critical point ($T\sim 10$-$15$ MeV in our
result). Beyond the critical point, $\protect\eta /s$ turns into a smooth
valley. The valley could disappear far away from the critical point. Similar
behavior is also seen in water shown in Fig. 4. We suspect these are general
features for first-order phase transitions.}
\end{figure}


Now let us focus on the $\mu \simeq m_{N}$ and $T<30$ MeV region in the $%
\eta /s$ plot. Two 3-D plots viewed from different angles are shown in Fig.
3(a) and 3(b) and a contour plot is shown in Fig. 3(c). One clearly sees
that $\eta /s$ maps out the nuclear gas-liquid phase transition shown in
Fig.1 by forming a valley tracing the nuclear gas-liquid phase transition
line in the $T$-$\mu $ plane. When the phase transition turns into a
crossover at larger $T$, the valley also gradually disappears at around 30
MeV. This result is encouraging.

However, even though the gross features of the phase transition are mapped
out by the $\eta /s$ valley nicely, some details are not correct. First,
since the density is discontinuous across the first-order phase transition, $%
\eta $, $s$, and $\eta /s$ are likely to be discontinues across the phase
transition as observed in H$_{2}$O, He and N systems in Ref. \cite%
{Csernai:2006zz}. This discontinuity, which defines the critical chemical
potential $\mu _{c}$, should lie at the bottom of the $\eta /s$ valley.
Second, the position of $\mu _{c}$ suggested by our result is not correct.
Near $T=0$, one expects $\mu _{c}\simeq m_{N}-\left\langle B\right\rangle $,
where $\left\langle B\right\rangle $ is the binding energy per nucleon, but
we have $\mu _{c}>m_{N}$.

It is quite obviously that our free particle treatment of $s$ is very poor
near the phase transition. However, we have not pursued other treatments
like the mean field approximation in this work because it is known that the
approximation is insufficient when $\left\vert k_{F}a\right\vert \gg 1$. For
the same reason, the computation of $\eta $ using the Boltzmann equation
might not be justified near the bottom the valley even though the mean free
path is still bigger than the range of potential ($\sim 1$ fm). However, the
valley of $\eta $ is located at $\mu <m_{N}$ near $T=0$; thus, it is
possible that after reliable $s$ is used, $\mu _{c}$\ for $\eta /s$ will be
in the correct position. Furthermore, the regime in Fig. 3 is completely
dominated by the nucleon degree of freedom ($\eta /s$ hardly changes with
the thermal pions completely ignored). This simplifies the problem
significantly and makes the system exhibit universal properties shared by
dilute fermionic systems with large scattering lengths such as cold atoms
tuned to be near a Feshbach resonance.

As mentioned above, it was observed that below the critical pressure, $\eta
/s$ has a discontinuity at the critical temperature for H$_{2}$O, He and N,
and above the critical pressure, $\eta /s$ has a smooth minimum near the
crossover temperature (defined as the temperature where the density changes
rapidly) \cite{Csernai:2006zz}. For QCD at $\mu =0$, $\eta /s$ also has a
valley near the crossover temperature \cite{Chen:2006ig,Csernai:2006zz}. But
there is no evidence yet to show the valley is smooth or has a
discontinuity. 

\begin{figure}[tbp]
\begin{center}
\includegraphics[height=8cm]{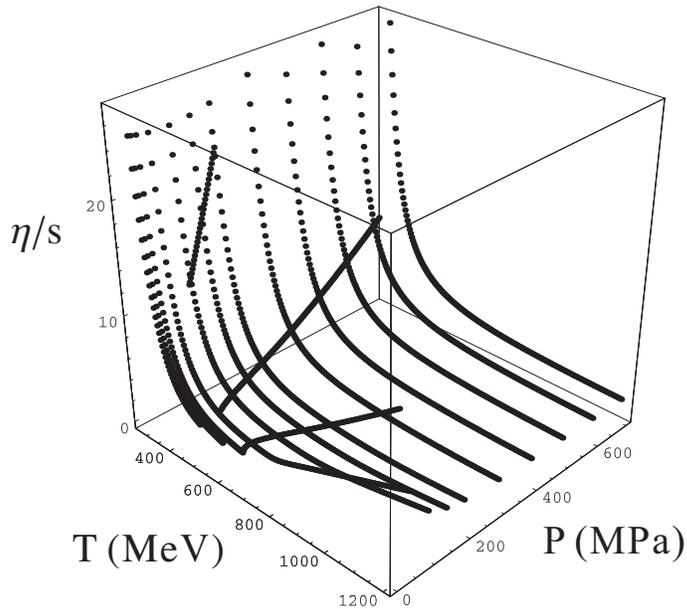}
\end{center}
\caption{$\protect\eta /s$ of water shown as a function of temperature and
pressure. Below the critical pressure 22.06 MPa, $\protect\eta /s$ has a
discontinuity at the bottom of the valley. Above the critical pressure, the
valley becomes smooth and the bottom (minimum) of the valley moves toward
the large $T$ and large $P$ direction. In a limited range of $T$, $\protect%
\eta /s$ could look like a monotonic function without a valley far away from
the critical pressure. The similar behavior is seen in all the materials
with data available in \protect\cite{webbook,codata}, including Ar, CO, CO$%
_{2}$, H, He, H$_{2}$S, Kr, N, NH$_{3}$, Ne, O, and Xe.}
\end{figure}


We suspect $\eta /s$ should be smooth in a crossover and should have a
discontinuity across a first-order phase transition.\emph{\ }If this is
correct, then in nuclear gas-liquid transition, there should be a
discontinuity looking like a fault in the bottom of the $\eta /s$ valley.
The fault would lie on top of the phase transition line and end at the
critical point where the first-order phase transition turns into a
crossover. If we look at Fig. 3(c), we would conclude that the critical
point is at $T\sim 10$-$15$ MeV, agreeing with $7$-$16$ MeV from
experimental extractions \cite{Natowitz:2001cq,Elliott:2002dk,Moretto:2005da}%
. Near the critical point, a smooth $\eta /s$ valley is seen in the
crossover (like the confinement-deconfinement crossover of QCD at $\mu =0$);
however, the valley could disappear far away from the critical point. We
suspect these are general features of first-order phase transitions. Indeed,
this behavior is seen in all the materials with data available in the NIST
and CODATA websites \cite{webbook,codata}, including Ar, CO, CO$_{2}$, H,
He, H$_{2}$O, H$_{2}$S, Kr, N, NH$_{3}$, Ne, O, and Xe. As an example, we
plot the $\eta /s$ of H$_{2}$O as a function of $T$ and $P$ in Fig. 4. Below
the critical pressure 22.06 MPa, $\eta /s$ has a discontinuity at the bottom
of the valley. Above the critical pressure, the valley becomes smooth and
the bottom (minimum) of the valley moves toward the large $T$ and large $P$
direction. In a limited range of $T$, $\eta /s$ could look like a monotonic
function without a valley far above the critical pressure. Thus, one might
use $\eta /s$ measurements to identify the first-order phase transition and
the critical point (see Refs. \cite{Csernai:2006zz,Lacey:2006bc} for a
similar point of view).

Also, for all the materials listed above except CO, the $\eta /s$ has a
positive jump across the phase transition line from the gas to the liquid
phase as shown in Fig. 4. For CO, the jump is positive at smaller pressure
but becomes small and negative at higher pressure.\ It is interesting to
recheck wether CO really is an anomaly. However, even without considering
CO, the sign of the $\eta /s$ \ jump for first order phase transitions might
not be universal, either. For QCD in the limit of a large number of colors,
the jump is negative from the low to high temperature phases \cite%
{Csernai:2006zz}.

\section{Conclusion}

We have computed the shear viscosity of QCD in the hadronic phase by the
coupled Boltzmann equations of pions and nucleons in low temperatures and
low baryon number densities. The ratio $\eta /s$ maps out the nuclear
gas-liquid phase transition by forming a valley tracing the phase transition
line in the temperature-chemical potential plane. When the phase transition
turns into a crossover, the $\eta /s$ valley also gradually disappears. We
suspect the general feature for a first-order phase transition is that $\eta
/s$ has a discontinuity in the bottom of the $\eta /s$ valley. The
discontinuity coincides with the phase transition line and ends at the
critical point. Beyond the critical point, a smooth $\eta /s$ valley is seen
on the crossover side. However, the valley could disappear far away from the
critical point. The $\eta /s$ measurements might provide an alternative to
identify the critical points.

\section{Acknowledgements}

We thank Dick Furnstahl for pointing out Refs. \cite%
{Natowitz:2001cq,Elliott:2002dk,Moretto:2005da} to us. This work was
supported by the NSC and NCTS of Taiwan, ROC.


\end{document}